# StegBlocks: ensuring perfect undetectability of network steganography

Wojciech Frączek, Krzysztof Szczypiorski
Institute of Telecommunications
Warsaw University of Technology
Warsaw, Poland
e-mail: wfraczek@gmail.com, ksz@tele.pw.edu.pl

*Abstract* — **The paper presents StegBlocks, which defines a new concept for performing undetectable hidden communication. StegBlocks is a general approach for constructing methods of network steganography. In StegBlocks, one has to determine objects with defined properties which will be used to transfer hidden messages. The objects are dependent on a specific network protocol (or application) used as a carrier for a given network steganography method. Moreover, the paper presents the approach to perfect undetectability of network steganography, which was developed based on the rules of undetectability for general steganography. The approach to undetectability of network steganography was used to show the possibility of developing perfectly undetectable network steganography methods using the StegBlocks concept.**

## I. INTRODUCTION

Network steganography incorporates all information hiding techniques that may be used to exchange secret data in telecommunication networks. This term was originally introduced by Szczypiorski in 2003 [1]. In network steganography, hidden communication is facilitated by methods which utilise network protocols or relationships between them as a carrier [2]. It is important to note that an observer (a third party) who is not aware that they are using the steganographic method is also not aware of the exchange of hidden data.

Network steganography communication can be characterised by four features: bandwidth, undetectability, robustness, and cost. The first three features, introduced by Fridrich in 1998 [3], are often presented as vertices of a triangle in order to show the interdependence among them. For example, the higher the required bandwidth is, the more difficult it is to achieve high undetectability and robustness. This interdependence causes the need for a trade-off between the three features when a new steganographic system is designed.

The last feature mentioned – cost – indicates the degradation of the carrier caused by the insertion of the secret data procedure [4]. It is important to note that the cost depends on each particular carrier and could be expressed in many ways, e.g. in-creased delay of packets, increased bit error rate, etc.

It should be emphasised that the most desirable characteristic of network steganography communication is usually undetectability [5].

There are many known network steganography methods. However, most of them can be easily detected if the algorithm of the method is known to the third party observer. Despite this fact, many authors of steganographic methods claim that the large number of possible ways of inserting hidden data into network communication makes detection of their methods extremely hard. The main focus of the analysis of undetectability of network steganographic methods is to show that the steganographic cost is very low. There are very few examples of the analysis of undetectability based on information theory or computation power, making it very difficult to assess the negative impact network steganography may have on our lives.

To address the above-mentioned problem, in this paper we propose the StegBlocks concept, which was designed to enable undetectable network steganography communication. StegBlocks is not a method based on a single network protocol. It defines an algorithm of secret communication and requirements for a network protocol (carrier) which could be used to create the steganographic method.

StegBlocks is used in this paper to show the possibility of constructing a network steganography method that is perfectly undetectable. The perfect undetectability is based on the information-theoretic approach, which has been described for general steganography in several publications [6-8]. We adjusted this approach to cover specific features of network steganography.

In this article we focused on perfect undetectability, which is parallel to the perfect security of cryptographic ciphers.

The rest of the paper is structured as follows. Section 2 introduces the concept of StegBlocks together with examples of methods built based on it. Section 3 describes the approach to the perfect undetectability of network steganography communication. Section 4 presents the proof of the possibility of creating a perfectly undetectable method based on StegBlocks. Finally, Section 5 concludes our work.

## II. STEGBLOCKS

### A. An idea

StegBlocks works using carriers (network protocols), for which elements (called objects) with various identifiers can be defined and the number of possible identifiers is limited. Once the objects are defined, the steganographic method created using StegBlocks works in the following way:

1. Among $n$ defined objects (objects with $n$ different identifiers), choose $k$ to be used as the steganographic key. The identifiers of the chosen objects have to be known to both the sender and the receiver of the hidden data.
2. Objects are merged into sequences (called blocks). A block is defined as a minimal-length subsequence in which there is at least one of each $k$ steganographic key object. The beginning of a block has to be defined, e.g. the end of one block determines the beginning of the next block.
3. Each block has an assigned value, which is the last $x$ bits of the number of objects in this block. The number of objects in the block includes all objects, not only objects which belong to the steganographic key.
4. The values of the blocks indicate bits of the hidden message. If the value of the block which is to be transmitted is not equal to the bits of the hidden message, then the sender has to change the value of the block.

The example of StegBlocks for four objects {1, 2, 3, 4}, from which 1 and 3 constitute steganographic keys, is presented in Fig. 1. The sender sends objects which form blocks based on the steganographic key. The first block consists of three objects {1, 4, 3}, because it is the minimal-length sequence of an object, which contains all of the objects of the steganographic key starting from the beginning of the transmission. The subsequent blocks are determined in the same way. Additionally, in the communication presented in Fig. 1 it was assumed that the last 1 bit of the number of objects in a block specifies the value of the block. This means that the value of the first block is 1 (the block consists of 3 ($11_2$) objects), the value of the second block is 0 (the block consists of 6 ($110_2$) objects), etc. The values of the blocks indicate the value of the hidden message, so in the presented example the hidden bits 1010 are transmitted. The receiver has to know the steganographic key in order to extract the secret message.

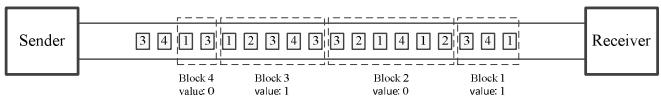

**Fig. 1 Example of StegBlocks**

The key element of StegBlocks is changing the values of the blocks in order to match them with the bits of the hidden message. This change can be made in three ways:
- control of the current object,
- control of the buffer,
- control of the entire communication.

Control of the current object is the approach in which, before sending the next objects, it is checked whether an object will complete the block. If the next object completes the block with a value which does not match the bits of the secret message, the sender suspends the sending of this object. Then the sender waits for next objects and the suspended object is sent when it is possible to complete the block with the required value. It could happen that the sender will have to suspend more than one object.

Control of the buffer is the approach in which all objects are placed in a buffer of size s (s is the number of objects, $s \geq 2$) before sending. In comparison with the control of the current object, in this case the sender knows s subsequent objects which are to be sent, allowing for more possibilities of changing the order of the objects.

Control of the entire communication is the approach in which the sender has access to all objects before the start of the communication. This approach allows for both the orders of any objects and the objects that are to be sent to be changed. In particular, it is possible to generate many possible sequences of objects to be sent in an overt channel and choose one, whose objects form blocks with values equal to bits of the secret message.

### B. Examples of network steganography methods created based on StegBlocks

This section presents two examples of network steganography methods based on StegBlocks. One of the methods uses TCP (Transmission Control Protocol) and the other method uses SCTP (Stream Control Transmission Protocol).

#### 1) StegBlocks TCP

The StegBlocks TCP method uses parallel TCP connections between two chosen hosts. The objects of this method are TCP segments transmitted within the parallel connection, and the identifiers of objects are the identifiers of the connection. The identifiers of connection (1, 2, ...) are assigned starting from the connection that was established first. A connection that was established earlier has a lower identifier than a connection established later. TCP segments within a given connection are objects with the same identifier. StegBlocks TCP works in the following way:

1. Two hosts, which are the sender and the receiver of the hidden data, have $n$ TCP connections established between them. Segments sent within a chosen connection are StegBlocks TCP objects with the same identifier. The number of unique identifiers of objects is equal to the number of used TCP connections.
2. Before the start of steganographic communication, the steganographic key is chosen ($k$ objects (TCP segments) with unique identifiers (TCP connections identifiers), which will determine the blocks).
3. A block is defined as the minimal-length sequence of an object, which contains at least one segment from each of the $k$ chosen in the previous point TCP connections. There are two ways of determining the start and end of the block. In the first one, the start of the block is determined by the end of the previous block. In the second one, the start of the block is determined by sending a specified number of segments from the start of the previous block. In both cases it could be assumed that the start of the first block is the start of the transmission.
4. The value of each block is the last $x$ bits of the number of TCP segments in this block. The number of segments in a block refers to all the segments in a block and not only segments within the chosen $k$ connections which determine the blocks.
5. The value of a chosen block determines the bits of the hidden message. If the value of a block is not equal to the bits of the hidden message, the sender has to appropriately modify the value of the block by changing the order of the segments, which are sent within parallel TCP connections.

The main problem with the presented method is the robustness of steganographic data transmission. Extracting the hidden message depends on the order of the reception of TCP segments. If one of the segments is significantly delayed or lost, it could influence the correctness of a received secret message. In order to avoid this issue, the sender has to wait to send the next segment of a message until they have received an acknowledgement of the reception of the previous segment. However, this solution may negatively affect the delays of an overt communication.

*2) StegBlocks SCTP*

The StegBlocks SCTP method works similarly to StegBlocks TCP, but in this case the problem of robustness is eliminated.

The StegBlocks SCTP method uses one SCTP connection with multiple streams. The objects of this method are chunks with data (DATA chunks) of the SCTP proto-col sent within established streams. The identifiers of objects are identifiers of the streams. Other aspects of the StegBlocks SCTP method are analogous to the Steg-Blocks TCP method.

The main advantage of the StegBlocks SCTP compared with StegBlocks TCP is the possibility of ensuring the robustness of the steganographic data transmission. In the case of StegBlocks SCTP, the receiver can determine the order of objects based on the TSN (Transmission Sequence Number) field, and not based on the time of the reception of specific objects (as in the case of StegBlocks TCP). The TSN is the sequence number for the DATA chunks, which is incremented and assigned to each chunk independently of the stream within which the chunk was sent. In the case of the delay or loss of a chunk, it can be retransmitted with the same TSN, which allows the correct reception of a secret message to be independent of the reception time of each object of the StegBlocks SCTP method.

### III. PERFECT UNDETECTABILITY OF NETWORK STEGANOGRAPHY

Perfect undetectability means that an adversary with unlimited computational power is not able to state if in a given overt communication there is also a hidden message transmitted. There are three popular definitions of perfect undetectability (based on information theory), which were introduced mainly for general steganography. All three definitions are derived from Shannon's definitions of the perfect security of a cryptographic system.

In 1998 Zöllner et al. presented a definition [6] in which steganography is undetectable if a hidden message is independent from a carrier and a steganogram (a carrier with a hidden message). The authors showed that a steganographic system is undetectable if the adversary is not able to compare a carrier and a steganogram. In order to achieve it, the adversary cannot know what a carrier without a hidden message looks like.

Cachin also proposed a definition of undetectable steganography in 1998 based on relative entropy (Kullback–Leibler divergence) [7]. According to this approach the steganographic system is undetectable if the relative entropy of the distribution of carriers without hidden data and the distribution of carriers with hidden data is equal to 0.

In 1999 Mittelholzer proposed a common definition of the security of steganography and watermarks [8]. According to this definition, a steganographic system is undetectable if it is not possible to distinguish a carrier and a carrier with hidden data. In order to measure this, the author suggested using an appropriate distortion measure, e.g. squared error distortion.

Among the presented known definitions of perfect undetectability of steganography, the most comprehensive one is the definition suggested by Cachin (Def. 1) [7]. The definitions proposed by Zöllner et al. [6] and Mittelholzer [8] contain necessary conditions, but are not sufficient for undetectable steganography [7].

**Def. 1.** (Perfect undetectability of a steganographic method according to [7]) A steganographic method is perfectly undetectable if:

$$D(P_C||P_S) = \sum_i P_C(i) \log \frac{P_C(i)}{P_S(i)} = 0, \qquad (1)$$

where:
$P_C$ is the distribution of carriers without hidden data;
$P_S$ is the distribution of carriers with hidden data;
$D(P_C||P_S)$ is the relative entropy of the distributions $P_C$ and $P_S$.

Additionally, the author in [7] suggests assuming that the adversary knows the hidden message that could be sent using steganography. Consequently, an insertion algorithm cannot depend on the distribution of the hidden message.

The presented definition is appropriate for steganography in the content, e.g. images. However in the case of network steganography it is suggested that three elements of network steganography communication be included in this definition:
- data units (e.g. packets, segments);
- transmission (time between sending/receiving data units);
- hidden message.

The mentioned elements were incorporated in the definition of the perfect undetectability of network steganography (Def. 2), which is based on Def. 1.

**Def. 2.** (Perfect undetectability of a network steganography method) A network steganography method is perfectly undetectable if all the following conditions are satisfied:
1. The conditional entropy of the steganogram and hidden message is equal to the entropy of the steganogram $(H(S|M) = H(S))$.
2. The relative entropy of the distribution of data units of a carrier without hidden data ($P_C$) and the distribution of a carrier with hidden data ($P_S$) is equal to 0 ($D(P_C||P_S) = 0$).
3. The relative entropy of the distribution of times between sending/receiving data units for a carrier without hidden data ($P_{C_t}$) and the distribution of times between sending/receiving data units for a carrier with hidden data ($P_{S_t}$) is equal to 0 $(D(P_{C_t} || P_{S_t}) = 0)$.

The presented definition of the perfect undetectability of a network steganography method will be used to prove that a steganographic method based on StegBlocks can be perfectly undetectable.

IV. PERFECT UNDETECTABILITY OF STEGBLOCKS

StegBlocks helps in the creation of perfectly undetectable network steganography methods. In order to achieve this, it is necessary to impose additional requirements for the method based on StegBlocks. The key requirement is use of the Vernam cipher to encrypt a hidden message before inserting it into a carrier.

**Theorem 1.** StegBlocks with the encryption of a hidden message using the Vernam cipher and the control of the entire communication allows for the creation of a network steganography method that is perfectly undetectable according to Def. 2.

**Proof of Theorem 1**
The proof uses a network steganography method based on StegBlocks with the following assumptions:
- $n$ unique identifiers of objects belonging to the set $\{1,2,3,...n\}$ are used;
- Objects are sent in groups and each group has n objects with all possible values of identifiers;
- Groups of objects are independent and the probability of the occurrence of each group of objects is the same $(1/n!)$;
- The start of a StegBlocks block is the first object sent within a given group;
- The last bit of the number of objects in the block is used as the value of the block;
- The probability of occurrence of a block with value 1 or 0 is equal to 1/2 (it is possible to achieve by appropriate choice of parameters $n$ and $k$ StegBlocks, e.g. n even and $k$ equal to 1);
- A change of the value of a block can be made by changing the order of single objects or changing the order of groups of objects (and thereby blocks of StegBlocks);
- A hidden message is encrypted using a one-time pad (the Vernam cipher) before inserting the message into a carrier.

Data units for the method with the presented assumptions are groups of $n$ objects.

The proof was divided into three parts corresponding to three conditions required to achieve perfect undetectability according to Def. 2:
- Knowledge of bits of the hidden message (condition 1: $H(S|M) = H(S)$);
- Distribution of data units (condition 2: $D(P_C||P_S) = 0$);
- Distribution of times between sending/receiving data units (condition 3: $(P_{C_t} || P_{S_t}) = 0$).

The theorem was proved by showing that all three conditions required for perfect undetectability of a network steganography method (according to Def. 2) are satisfied for the method based on StegBlocks with the above-mentioned assumptions.

*1) Knowledge of bits of hidden message*

In order to show that knowledge of bits of a hidden message does not allow for determining if a network steganography method based on StegBlocks is used, it is necessary to check the value of the conditional entropy of a steganogram and a hidden message. If the conditional entropy of a steganogram and a hidden message is equal to the entropy of a steganogram, then the knowledge of a hidden message does not allow for any information on the steganogram to be gained. This means that knowledge of a hidden message does not help in determining if network steganography has been used.

The condition $H(S|M) = H(S)$ is equivalent to condition $Pr(S = s|M = m) = Pr(S = s)$ for each $s \in S$ and $m \in M$ (s is a steganogram and $m$ is a hidden message). According to the assumptions regarding the considered network steganography method the probability of block of value 1 or 0 is equal to 1/2. Let $l$ be the length of the key of a one-time pad, and therefore at the same time the length of a secret message and a ciphertext. After using an the one-time pad $S = M \oplus K$, this becomes:

$$\begin{aligned} Pr(S = s|M = m) &= Pr(M \oplus K = s|M = m) \\ &= Pr(m \oplus K = s) \\ &= Pr(K = m \oplus s) \\ &= \frac{1}{2^l} \\ &= Pr(S = s). \end{aligned} \quad (2)$$

*2) Distribution of data units*

Each data unit of the considered network steganography method (group of $n$ objects) is a permutation of the $\{1,2,3, \ldots n\}$ set. According to the assumptions of the method, groups of objects are independent and the probability of the occurrence of each group is the same ($1/n!$). Let $\{e_0, e_1, e_2, \ldots e_{n!}\}$ be a set of all possible permutations of the $\{1,2,3, \ldots n\}$ set and let the values of blocks determined by the groups $e_i$, where $i$ is odd, be equal to 1, and the values of other blocks be 0. In addition, let $X = x_1, x_2, \ldots, x_w$ be a hidden message encrypted with the Vernam cipher. Then, assuring that values of blocks are equal to bits of encrypted hidden message can be achieved through the choice of subsequent groups of objects in the following way:

$$s_j = e_{(id_j + (x_j \oplus v_j)) \bmod n!} \quad (3)$$

where:
$id_j$ is an identifier of the j-th group of objects, which would be sent if steganography were not used;
$v_j$ is a value of a block determined by the j-th group of objects, which would be sent if steganography were not used;
$s_j$ is the j-th element of the steganogram (group of objects).

Due to the fact that $x_j$ are the subsequent bits of a hidden message encrypted with the Vernam cipher, the applied way of assuring that values of blocks are equal to bits of an encrypted hidden message is not changing the probability distribution of data units, so the relationship $D(P_N || P_S) = 0$ is true.

*3) Distribution of times between sending/receiving data units*

Two factors influence the times between sending/receiving data units in the case of using a network steganography method based on StegBlocks:
- delays caused by inserting secret message into a carrier,
- ensuring error-free steganographic communication in the case of additional delays and losses during transmission.

It is possible to eliminate the delays caused by inserting a secret message into a carrier by using the control of the entire communication as a way of changing values of blocks of StegBlocks. The determination of objects to be sent before the transmission actually starts prevents potential additional delays during connection if the order of the objects needs to be changed.

Delays and losses during transmission may influence the method of changing the order of objects based on StegBlocks, which could lead to the incorrect extraction of the hidden message by the receiver. In the case of the StegBlocks SCTP method this is not an issue, as the order of the objects is determined based on the TSN, not the actual time of their reception. Therefore, there is no need for an additional mechanism preventing the incorrect reception of steganographic data and there is no influence on the times between sending/receiving data units ($D(P_{N_t} || P_{S_t}) = 0$).

Ensuring the absolute undetectability of the StegBlocks TCP method requires an acceptance of potential errors in the steganographic transmission. If there are no additional actions related to ensuring error-free steganographic communication (e.g. waiting to send the next segment until acknowledgement of the reception of the previous segment has been received, as mentioned in Section 2), the distribution of the times between sending/receiving data units will not differ from the distribution for transmission without steganography ($D(P_{N_t} || P_{S_t}) = 0$).

Summing up, StegBlocks allows for the creation of a network steganography method for which the relative entropy of the distribution of times between sending/receiving data units for a carrier without a hidden message and the distribution of times between sending/receiving data units for a carrier with a hidden message is equal to 0.

The proofs of the satisfaction of all conditions of Def. 2 conclude the proof that StegBlocks allows for the creation of a network steganography method that is perfectly undetectable.

V. CONCLUSIONS

In this paper, a new concept for performing undetectable network steganography communication, called StegBlocks, was presented. StegBlocks is a framework for the creation of network steganography methods. It defines the way in which the methods work and at the same time allows for the creation of methods for various carriers (network protocols).

Moreover, the paper contains the definition of perfectly undetectable network steganography. The definition is based on the definitions for general steganography developed by Cachin [7]. We adjusted this definition to cover the specific features of network steganography.

Finally, the definition of perfect undetectability was used to show that it is possible to create a network steganography method based on StegBlocks which is perfectly undetectable (undetectable for the adversary with unlimited computational power).

Although the ideas presented in this paper may seem to have limited practical applications, their objective is to show the possibility of a more thorough approach to the analysis of the undetectability of network steganography than is currently used in most publications. Most of the publications related to network steganography do not focus enough on undetectability. We hope that this article will spur more research regarding the undetectability of network steganography.